\documentstyle[aps,prl,graphics]{revtex}

\newcommand{\be}{\begin{equation}}
\newcommand{\ee}{\end{equation}}
\newcommand{\bea}{\begin{eqnarray}}
\newcommand{\eea}{\end{eqnarray}}
\newcommand{\beb}{\begin{eqnarray*}}
\newcommand{\eeb}{\end{eqnarray*}}

\newcommand{\SrCu}{SrCu$_2$(BO$_3$)$_2$\ }
\begin{document}


\twocolumn[\hsize\textwidth\columnwidth\hsize\csname@twocolumnfalse\endcsname

\title{Magnetization Plateaus of \SrCu from a Chern-Simons Theory}

\author{G.~Misguich$^a$, Th.~Jolicoeur$^b$ and S.~M.~Girvin$^{c,d}$}
\address{
{\it (a)} Service de Physique Th\'eorique, CE Saclay,
91191 Gif-sur-Yvette, France\\
{\it (b)} D\'epartement de Physique, LPMC-ENS, 24 rue Lhomond, 
75005 Paris, France\\
{\it (c)} Department of Physics, Indiana University, Bloomington,
Indiana 47405-7105, USA\\
{\it (d)} Institute for Theoretical Physics, UCSB, Santa Barbara, CA
93106-4030}

\date{February 21, 2001}
\maketitle
\begin{abstract}
The     antiferromagnetic  Heisenberg  model     on   the   frustrated
Shastry-Sutherland lattice  is  studied by  a mapping   onto  spinless
fermions carrying one quantum of statistical flux.        
Using a  mean-field
approximation  these fermions populate  the   bands of a   generalized
Hofstadter  problem. Their filling leads to the magnetization curve.
For \SrCu we reproduce plateaus at 1/3 and 1/4 of the saturation moment      
and predict a new one at 1/2.               
Gaussian fluctuations of the gauge field 
are shown to be massive at these plateau values.
\end{abstract}


\vskip2pc]

\tighten\narrowtext

\begin{figure}
\begin{center}
\resizebox{7.5cm}{!}{\includegraphics{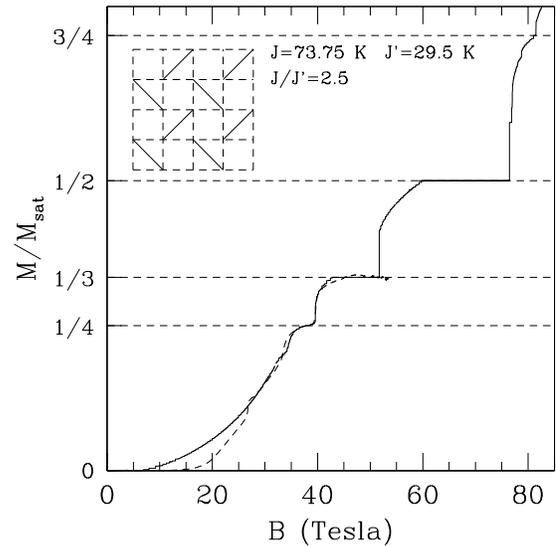}}
\end{center}
\caption{
Comparison  between the  magnetization  curve  of SrCu$_2$(BO$_3$)$_2$
measured by  Onizuka  {\it et  al.}  (dashed line) and the  mean-field
result (solid line). Inset:  Shastry-Sutherland lattice. The  exchange
interaction is $J$ on black links and $J'$ on the dotted ones.
}
\label{magexp}
\end{figure}

Two-dimensional     (2D)  quantum  spin  systems  that    do not order
magnetically at  zero temperature are  currently  a subject  of  great
theoretical       and    experimental     interest.     The   recently
discovered\cite{kage99} compound \SrCu is an antiferromagnet (AF) with
localized    spins   $S=\frac{1}{2}$,   a   gap  above    the   ground
state\cite{miyahara99} and  the unique property that its magnetization
curve has  plateaus at $1/3$, $1/4$, and  $1/8$ of the full saturation
moment\cite{onizuka00}.   The  spin system may  be   described by a 2D
Heisenberg model  on a square lattice  with exchange constant $J'$ and
additional  diagonal bonds $J$  on half of  the square plaquettes
(see inset of Fig.~\ref{magexp}).

This   lattice  was     studied  many  years   ago  by    Shastry  and
Sutherland\cite{ss81} who  noted  that there  is an  exact  eigenstate
which is obtained by putting singlets on all diagonal $J$ bonds.  This
eigenstate is  the ground state  for a wide  interval  of $J'/J$.  For
$J'/J$ smaller than $\sim 0.7$\cite{weihong99},  the system has dimer
long-range  order and  for  larger  $J'$  it  has  conventional N\'eel
long-range AF order.  There may be  additional phases in between 
such as a plaquette singlet   phase\cite{koga00,CB01,CMS01}  but they  
are apparently not realized in \SrCu where $J'/J$ is estimated to be 
smaller  than $0.65$\cite{miyahara00b,kbmu00}.  
This  dimer ground state explains the
spin gap as seen in  experiments.  However  the existence of  plateaus
has no immediate explanation since the simplicity  of the ground state
does not  extend to the excited states  of this  peculiar lattice.  In
this Letter we use a mean-field approximation to a Chern-Simons
(CS) field-theoretic   approach to quantum  magnets suggested
some time ago \cite{fradkin89,as89} 
and we obtain  an excellent quantitative fit of the magnetization curve  
(see Fig.~(\ref{magexp})) for realistic values of the exchange constants.  
As we discuss, this may be evidence for
unconventional  character of the plateau ground states.

Starting from the pure  dimer state, the   first excited state can  be
constructed by first breaking a singlet bond into a triplet state. Due
to the exchange $J'$, such a state will move  and we expect a dispersive
band of  triplets as low-lying states  at least for strong $J'$.

 However
due to the  peculiar triangular coupling  between the diagonal  bonds,
the hopping of the triplet is  forbidden at low orders in perturbation
theory\cite{miyahara99}.  As a  consequence, the triplet band is very
flat,   a   striking   fact    observed   by     neutron   scattering
experiments\cite{kageyamaNeutrons}.    Since  the  triplets   are very
massive particles,  it is natural to expect  that they can crystallize
at finite  density, and it has been  proposed  that the plateaus  are
Wigner crystals of triplets\cite{miyahara00,momoi00,momoi00b}.  There
exists a  spin model  which is derived from the  Shastry-Sutherland
Hamiltonian\cite{msku00} for  which the  plateaus are  demonstrated to
originate from such ordered states.  However in  this model there are
plateaus at $1/4$, $1/2$ and $3/4$ and  the overall shape  of
the magnetization curve is not in agreement with experiments.
A closely related physical picture is obtained
by  describing the magnetized triplets by hard-core bosons\cite{momoi00}.
Then the   repulsion  may favor charge-density   wave  states that are
amongst known insulating phases of the lattice Bose gas.

We take a different approach here, mapping the spin problem onto a hard-core
boson problem and then solving the hard core constraint exactly by a further
mapping onto spinless fermions coupled to a CS gauge 
field\cite{as89,ywg93,lrf94}.
Within a mean-field approximation the spin excitation gap that produces the
observed magnetization plateaus arises from some of the Landau level gaps
in the
integer quantum Hall effect for the fermions on a lattice.

The Hamiltonian for the AF Heisenberg
model on the Shastry-Sutherland lattice is:
\be
H =\sum_{\left\langle i,j\right\rangle}
J_{ij} \, {\vec S}_{i}\cdot {\vec S}_{j}
-B \sum_i S^z_i ,
\label{HamSpin}
\ee
where   $\vec  S_i$  are  spin-$\frac{1}{2}$  operators,  the exchange
couplings $J_{ij}$ are equal to  $J'$ when $i ,j$ are  nearest-neighbors
on the square  lattice and equal to J   when $i ,j$  are  related by a
diagonal  bond ($J, J' > 0$) and the  external  magnetic field $B$ is
applied along the z-axis.  We then map the spin operators to hard-core
boson operators~:
\bea
H&=&H_{xy} + H_z \, ,\\
H_{xy}&=&{1\over 2}\sum_{\left\langle i,j\right\rangle} J_{ij}
\left( b^{\dag}_i b_j + b^{\dag}_j b_i \right) ,\\
H_z&=&\sum_{\left\langle i,j\right\rangle} J^z_{ij}
(n_i -1/2)(n_j -1/2)-B\sum_i \left(n_i -1/2\right) \, ,
\label{HamBose}
\eea
where $n_i\equiv S^z_i +1/2$ is the occupation number of site $i$. The
bosons  are  then treated as   fermions  with an  attached  flux  tube
carrying one  flux quantum of  fictitious magnetic field.  This can be
formulated as an  {\em exact}  mapping between  the  spin problem  and
spinless  fermions interacting with  a  statistical CS
gauge field.  In a mean-field treatment the flux tubes are smeared out
into   a uniform  background  magnetic  field.    The flux per  square
plaquette $\phi$ is then tied to  the density of  fermions and thus to
the magnetization of the spin system~:

\be
\frac{\phi}{2\pi}       = \langle n \rangle 
                        = (\langle S^z \rangle +{1\over 2})
                        = M+{1\over 2}
\label{fluxdensity}
\ee

It is important to note that the {\it real} magnetic field $B$ applied
to the spins  acts as a chemical  potential  for the fermions  as seen
from  Eq.~(\ref{HamBose})   but does  not   contribute   to  the  {\em
statistical}   flux.    We     treat   the  Ising    term   $H_z$   in
Eq.(\ref{HamBose}) by a simple mean-field  decoupling so it becomes  a
simple function of the magnetization. The kinetic energy term $H_{xy}$
leads to    a Hofstadter\cite{hofstadter76} problem    for   fermions  
hopping on the Shastry-Sutherland lattice. We use a flux  attachment
choice  leading  to  flux   $\phi/2$  on  each  triangular
plaquette.      The  one-body    problem    from    $H_{xy}$   can  be
straightforwardly analyzed for rational values $\phi=2\pi (p/q)$.

\begin{figure}
\begin{center}
\resizebox{!}{8.4cm}{\includegraphics{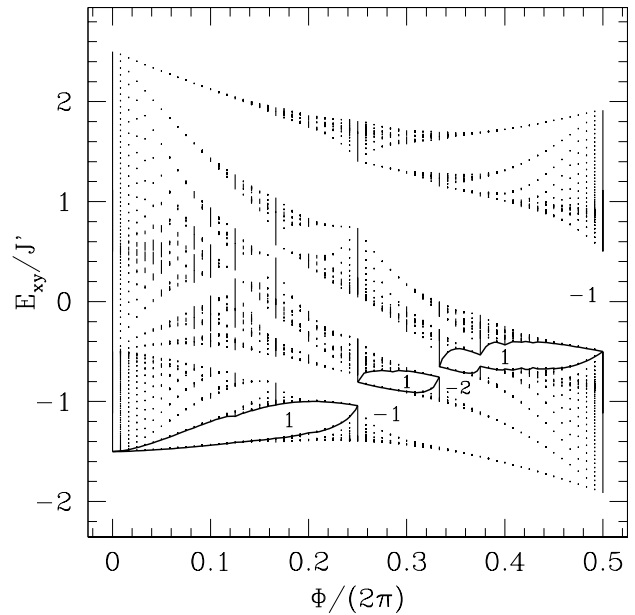}}
\end{center}
\caption{Hofstadter spectrum for the Shastry-Sutherland lattice
at  $J=J'=1$.
Vertical lines mark the energy bands as a function of the       
statistical flux $\phi$ per square plaquette.                   
Hall conductances $\sigma_{xy}$ (TKNN integers) are indicated   
for the regions of the spectrum which are explored in the       
magnetization process from                                      
$M/M_{sat}=-1$ ($\phi=0$) to $M/M_{sat}=0$ ($\phi=\pi$).
}
\label{sp1}
\end{figure}

\begin{figure}
\begin{center}
\resizebox{8cm}{!}{\includegraphics{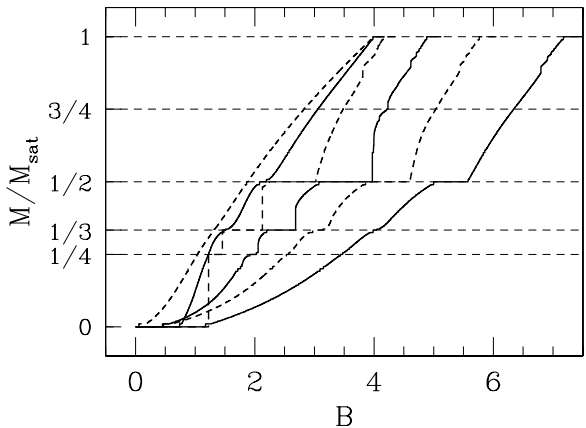}}
\end{center}
\caption{
Magnetization curves for the $J-J'$ Shastry-Sutherland model
(uniform mean-field). From left to right
$J=0$ (dashed) , 0.75 (full), 1.5 (dashed), 2.5 (full), 3.5 (dashed) and  5
(full). All these results are obtained for $J'=1$.}
\label{magss}
\end{figure}

\begin{figure}
\begin{center}
\resizebox{7.6cm}{!}{\includegraphics{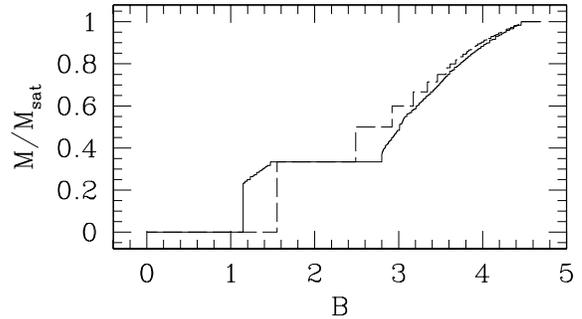}}
\end{center}
\caption{
Magnetization curve of the triangular antiferromagnet.
Full line: mean-field approximation with three-sublattices.
Dashed line: uniform mean-field.}
\label{magtri}
\end{figure}

\begin{figure}
\begin{center}
\resizebox{7.6cm}{!}{\includegraphics{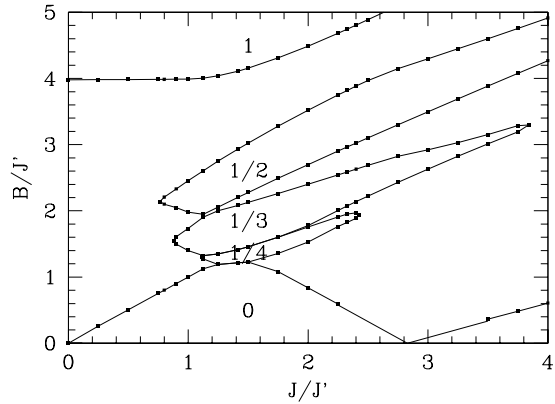}}
\end{center}
\caption{
Width of the magnetization plateaus as a function of $J/J'$
for $M/M_{\rm sat}=0, \frac{1}{2}, \frac{1}{3}$ and $\frac{1}{4}$. 
Additional plateaus at fractions $\frac{1}{n}$
for $n\ge5$ also exist in the vicinity of $J/J'\simeq1.5$.}
\label{plat}
\end{figure}

Fixing  the magnetization $M$  gives  us the flux   and the number  of
fermions through Eq.(\ref{fluxdensity}).  For  this value of $\phi$ we
compute  the band spectrum of $H_{xy}$ and fill the bands  
with the available fermions. The energy of the filled bands leads to a
first contribution $E_{xy} (M)$.   We then  add the contribution  from
the  Ising interaction $H_z$ to obtain  the  total energy $E(M)$.  The
magnetization is   obtained  as  a  function  of  $B$   by  minimizing
$E(M)-BM$.   The   Hofstadter   diagram  for    $J=J'$  is  given   in
Fig.~\ref{sp1} where
the lowest curve marks the Fermi level (highest occupied state) 
and the upper one marks the lowest unoccupied level.            
Jumps  of the
Fermi energy as a function of $M$ leads  to discontinuity of the slope
of the function  $E(M)$.  These jumps corresponds  to plateaus  in the
magnetization curve.  The structure  of the Hofstadter  butterfly thus
reflects itself in the appearance of plateaus.
The effects of geometry are encoded in  the Hofstadter spectrum, whose 
complexity arises from the diffraction of the cyclotron orbits by
the lattice.
When $J$ is set to zero, the model reduces to the square lattice which
has N\'eel  long-range order.  The  wavefunctions  obtained in  the 
spatially uniform CS mean-field   approximation certainly   do not
encompass  this physics.  However
the  magnetization curve obtained from  this approach is qualitatively
similar to that of the ordered  system~: it is  featureless all the way
to full saturation (see the $J=0$ curve in Fig.~\ref{magss}).   
If we consider the   case  of the  triangular  lattice, the  uniform
mean-field  leads to curve  (a) in  Fig.~\ref{magtri}.  In addition to
the zero-field gap found by Yang, Warman and Girvin\cite{ywg93}, there
are many  plateaus in (a). This is unrealistic~: the triangular lattice
spin system  is known to be long-range ordered and hence gapless and  
its magnetization process shows only a single plateau\cite{refmagtri} at
$M/M_{sat}=1/3$.  There is a way to improve  these results by allowing
the mean-field to have  a  three-sublattice structure: one  introduces
three fermion densities $n_A,  n_B, n_C$ and  numerically 
searches  for a
self-consistent   solution where  the  flux  $\phi_\alpha$ matches the
density  $n_\alpha$ on each  sublattice.  For  $M=0$ we  find that the
self-consistent   solution   remains uniform.   However,  for non-zero
magnetization the  translation  symmetry  is broken.   This non-uniform
mean-field   solution  leads   to     a  magnetization   curve   shown 
in Fig.~\ref{magtri}  which  is much  closer to the truth albeit  the
zero-field ground  state remains unrealistic (N\'eel long-ranged order
is absent). The  $1/3$ plateau  has a semiclassical origin~:
It is easily understood in the Ising limit  $J^z>>J$ limit      
and survives up to the isotropic point $J^z=J$\cite{refmagtri}.
Our calculation reproduces these  features and we find
the three sublattice magnetizations 
$n_A=n_B=0.922$ and $n_C=0.155$.
We have explored  the magnetization process of the  Shastry-Sutherland
lattice as a function of the ratio $J/J'$. The curves $M(B)$ are drawn
in Fig.~\ref{magss}.  For small $J$ we are close to the square lattice
result, i.e. a smooth  curve.  Increasing  $J$   we  observe  plateaus
developing,  the more complex  structure being  in  the regime  $J/J'
\approx  2-3$.
 For very large $J$ the curves are again simple.  
There remain only two plateaus  at $0$ and $1/2$. 
The plateau  at zero field is due to the fact that for large $J$  
the tight-binding bands separates  into   two groups with very   small
dispersion and  there is essentially  a  huge  gap in  the  Hofstadter
spectrum that does  not  depend much upon  flux. 
For $J'=0$ and an appropriate choice of gauge our wavefunction exactly
reproduces the dimer limit.
In  the  intermediate
regime, the  curve  depends  on  the  details of  the  spectrum.   The
plateaus  have   a finite  domain  of   stability which  is   given in
Fig.~\ref{plat}.
To reproduce the  qualitative shape of the experimental  magnetization
curve for \SrCu  we find that  $J'= 29.5$K and $J=74$K, the  resulting
fit is shown   in Fig.~\ref{magexp}. While the  zero-field  gap is not
well reproduced, we  find 13K instead of 34K,  
the field strengths at which the plateaus occur and the roundings close
to the plateaus are in very good agreement  with  the experiment.  
The values of  $J, J'$  are  reasonably  
close to  the recent estimates\cite{kbmu00}
from  neutron   scattering data $J'= 43 $K and $J= 71 $K.  
These values are not precisely known   since there  is  some  amount  
of  three-dimensional
dispersion of the low-lying triplet that should be taken into account.
We  predict prominent plateaus at $1/3$ and $1/2$ and nothing else until
full saturation.  Note also  that at $1/4$ there is in fact an avoided
plateau~: the value of $J'/J$ is just outside the  range of stability of
the $1/4$ plateau on  Fig.~\ref{plat}. The result of CS  mean-field
theory is quite different  from  the  semiclassical analysis of  the
effective Bose gas\cite{momoi00b} and from the exactly soluble spin
model\cite{msku00}.

As can be seen in Fig.~\ref{plat} the  spin gap (plateau at
$M/M_{\rm  sat}=0$) opens at   $J/J'=0$  in this  approach instead  of
opening at the correct  critical coupling $J/J'\sim1.5$.  This feature
comes from  the fact   that  the N\'eel  state of  the  square-lattice
antiferromagnet is  not  correctly  described  in  the  {\em  uniform}
mean-field  approximation, as mentioned above   in   the case of   the
triangular-lattice antiferromagnet.  This   is corrected by  computing
non-uniform   solutions  of     the   mean-field  CS    approach  with
two-sublattices\cite{lrf94}.

\null    From the good fit in Fig.~\ref{magexp} we deduce that
the CS mean-field theory works better for magnetized  than unmagnetized
states.   This is because
the initial mapping to hard core bosons selects out a
preferred spin quantization axis which, together with
the mean-field treatment of the Ising term, obscures the SU(2)
symmetry in the underlying spin problem.  
The Zeeman energy reduces the SU(2) to the U(1) symmetry associated
with conservation of $S^z$, which in the language of the fermions is
simply particle number conservation.  We note that the physical 
magnetic field breaks T symmetry in the orbital part of a Hubbard model
description of the system but has no effect in a pure spin model
description other than to introduce a Zeeman term.  However our mean
field solution does break T symmetry.  The increased size of the
`magnetic' unit cell (due to the CS flux) gives an integer number of
particles per unit cell as required for the existence of a 
gap\cite{oshikawa00} even though there is no translation symmetry
breaking in the spin density.

It  is  important to check  whether the  plateau states  are robust to
fluctuations beyond mean-field.  
Following an idea of Fradkin \cite{fradkin1990}  
Yang  {\it et al.}\cite{ywg93} showed
that Gaussian fluctuations of the CS  gauge field are massive provided
that the TKNN\cite{tknn82} integer describing the  quantized Hall 
coefficient  of the fermions on the frustrated   
lattice differs from the continuum 
value of {\em unity}. We have computed the TKNN integers by following 
the evolution of gaps from the square lattice case where the TKNNs are 
given by a Diophantine equation.  None of the plateaus is suppressed by 
fluctuations since we find $\sigma=-3$, $-2$ and $-1$ respectively 
for the plateaus at  $\frac{1}{4}$, $\frac{1}{3}$ and $\frac{1}{2}$ 
(see Fig.~\ref{sp1}). At $M/M_{sat}=0$ we have  $\sigma=-1$ (resp $0$) 
for $J/J'<\sqrt(2)$ (resp $J/J'>\sqrt(2)$).

A consequence of these non-trivial quantized Hall coefficients for the
fermions is that the spin state is chiral and exhibits a `spin quantum
Hall effect'\cite{HaldaneandArovas}. Whether this new physics 
is actually occurring
in \SrCu or is an unrealistic feature of  our mean field approximation
remains to be seen. 
One test of our model is the prediction of a strong plateau
in the magnetization beginning at 60T, a field which is within reach of
modern pulsed magnets.


The authors are grateful to L. Balents, E.~S.~Fradkin, M.~Kohmoto, 
S. Sachdev, G.~Semenoff, K.~Totsuka and K.~Ueda for useful discussions. 
SMG is supported by NSF DMR-0087133 and the ITP at UCSB under PHY94-07194.


\end{document}